\documentclass[conference]{IEEEtran}

\usepackage{amsmath,amsfonts}
\usepackage{graphicx}
\usepackage{booktabs} % For professional-looking tables
\usepackage{siunitx}  % For consistent unit formatting (e.g., \SI{80}{\milli\kelvin})

\usepackage{url}
\usepackage{textcomp}
\usepackage{subfig} % For subfigures, if needed
\usepackage{algorithmic} % For algorithms, if needed
\usepackage{balance} % To balance the columns on the last page
\usepackage{comment}
\usepackage[backend=biber, style=ieee]{biblatex}
\begin{document}

% --- TITLE ---
\title{CCAT: Mod-Cam Cryogenic Performance and its Impact on 280 GHz KID Array Noise}
\author{
    \hspace*{-0.5cm}% Manually shifts the entire block left by 0.5cm
    \begin{tabular}{@{}c@{}} % Use a single centered tabular environment
        % --- Author List ---
        Lawrence T. Lin$^{1}$, Eve M. Vavagiakis$^{2,1}$, Jason E. Austermann$^{3}$, James R. Burgoyne$^{4}$, Scott Chapman$^{5}$, Steve K. Choi$^{6}$, \\
        Abigail T. Crites$^{1}$, Cody J. Duell$^{1}$, Rodrigo G. Freundt$^{7}$, Eliza Gazda$^{6}$, Christopher Groppi$^{8}$, Anthony I. Huber$^{4}$, \\
        Zachary B. Huber$^{1}$, Johannes Hubmayr$^{3}$, Ben Keller$^{1}$, Philip Mauskopf$^{8}$, Alicia Middleton$^{1}$, Michael D. Niemack$^{1,7}$, \\
        Darshan A. Patel$^{1}$, Cody Roberson$^{8}$, Adrian K. Sinclair$^{9}$, Ema Smith$^{1}$, Anna Vaskuri$^{3}$, Benjamin J. Vaughan$^{1}$, \\
        Samantha Walker$^{1}$, Yi Wang$^{2}$, Yuhan Wang$^{1}$, Jordan Wheeler$^{3}$, Ruixuan(Matt) Xie$^{4}$
        \\[2.0ex] % Adds a little extra space between authors and affiliations

        % --- Affiliation List (Manually setting font for each line) ---
        \small $^{1}$Department of Physics, Cornell University, Ithaca, NY 14853, USA \\
        \small $^{2}$Department of Physics, Duke University, Durham, NC 27710, USA \\
        \small $^{3}$Quantum Sensors Division, National Institute of Standards and Technology, Boulder, CO 80305, USA \\
        \small $^{4}$Department of Physics and Astronomy, University of British Columbia, Vancouver, BC, Canada \\
        \small $^{5}$Department of Physics and Atmospheric Science, Dalhousie University, Halifax, NS, Canada \\
        \small $^{6}$Department of Physics and Astronomy, University of California, Riverside, CA 92521, USA \\
        \small $^{7}$Department of Astronomy, Cornell University, Ithaca, NY 14853, USA \\
        \small $^{8}$School of Earth and Space Exploration, Arizona State University, Tempe, AZ 85287, USA \\
        \small $^{9}$NASA Goddard Space Flight Center, Greenbelt, MD 20771, USA \\
        
    \end{tabular}
}

% This command generates the title, author, and abstract section.
\maketitle

% --- ABSTRACT ---
\begin{abstract}

The CCAT Observatory’s Fred Young Submillimeter Telescope (FYST) is designed to observe submillimeter astronomical signals with high precision, using receivers fielding state-of-the-art kinetic inductance detector (KID) arrays. Mod-Cam, a first-light instrument for FYST, serves as a testbed for instrument module characterization, including detailed evaluation of thermal behavior under operating conditions prior to deploying modules in the larger Prime-Cam instrument. Prime-Cam is a first generation multi-band, wide-field camera for FYST, designed to field up to seven instrument modules and provide unprecedented sensitivity across a broad frequency range.

We present results from two key laboratory characterizations: an ``optically open" cooldown to validate the overall thermal performance of the cryostat, and a ``cold load" cooldown to measure the effect of focal plane temperature stability on detector noise. During the optically open test, we achieved stable base temperatures of 1.5 K on the 1 K stage and 85 mK at the detector stage. In the cold load configuration, we measured a detector focal plane RMS temperature stability of 3.2e-5 K. From this stability measurement, we demonstrate that the equivalent power from focal plane thermal fluctuations is only 0.0040\% of a 5pW incident photon power for aluminum detectors and 0.0023\% titanium-nitride detectors, a negligible level for CCAT science goals. This highlights the success of the cryogenic system design and thermal management.
\end{abstract}

% --- KEYWORDS ---

% --- INTRODUCTION ---
\section{Introduction}
\IEEEPARstart{T}{he} next generation of millimeter and submillimeter surveys will require instruments with significantly improved sensitivity and mapping speeds to address key open questions in cosmology and astrophysics. The Fred Young Submillimeter Telescope (FYST) is a 6-meter aperture, crossed-Dragone telescope being constructed for the CCAT Observatory at an altitude of 5600 meters on Cerro Chajnantor in the Chilean Atacama Desert \cite{huber2024ccatprimecamopticsoverview}. Its large field of view is designed to provide enhanced mapping speeds in the submillimeter \cite{Niemack_2016}.

The advantages offered by the design of FYST will be utilized by the Prime-Cam receiver. Prime-Cam will field seven independent instrument modules to observe between 220 and 850 GHz, adapted from the Simons Observatory Large Aperture Telescope Receiver optics tube design \cite{Zhu_2021,sierra2024simonsobservatorypredeploymentperformance}, including two spectroscopically-enabled line intensity mapping imaging spectrometers \cite{Vavagiakis_2018} \cite{Freundt_2024}. The instrument modules will be populated with large format arrays of superconducting microwave kinetic inductance detectors (KIDs), with modules containing tens of thousands of detectors. For example, the 280 GHz module has 10,350 polarization-sensitive KIDs , and the full Prime-Cam instrument is designed to accommodate over 100,000 detectors in total. KIDs offer inherent frequency domain multiplexing (FDM) capabilities, which dramatically simplifies the readout method for thousands of pixels on a single coaxial line, reducing the complexity of cryogenic components \cite{Zmuidzinas_2012}.

As the single-module precursor to Prime-Cam, Mod-Cam is the designated first-light and commissioning instrument for FYST. To date, its primary function has been to serve as a critical in-lab testbed for the 280 GHz module. After Prime-Cam is deployed, Mod-Cam will continue its role as a testbed for future instrument modules. To enable complete testing of instrument modules and thermally sensitive detectors in Mod-Cam, it is crucial to understand the system's thermal stability. As KIDs are sensitive to their operating temperature \cite{tempKIDS}, any fluctuations in the cryogenic environment will induce a signal that can increase the thermal noise floor and degrade the instrument sensitivity.

The primary objectives of this paper are to describe %to the details surrounding 
the thermal performance of the Mod-Cam testbed; to measure the thermal responsivity ($\mathcal{R}_T$) of both aluminum (Al) and titanium-nitride (TiN) KIDs; to quantify the temperature stability of the detector stage; to calculate the equivalent optical power noise resulting from this thermal instability and demonstrate that it is a negligible contributor to the overall detector noise budget.

\section{Kinetic Inductance Detector Principles}
\begin{figure}[!t]
\centering
% User should replace 'fig1.png' with an actual file.
\includegraphics[width=0.6\columnwidth]{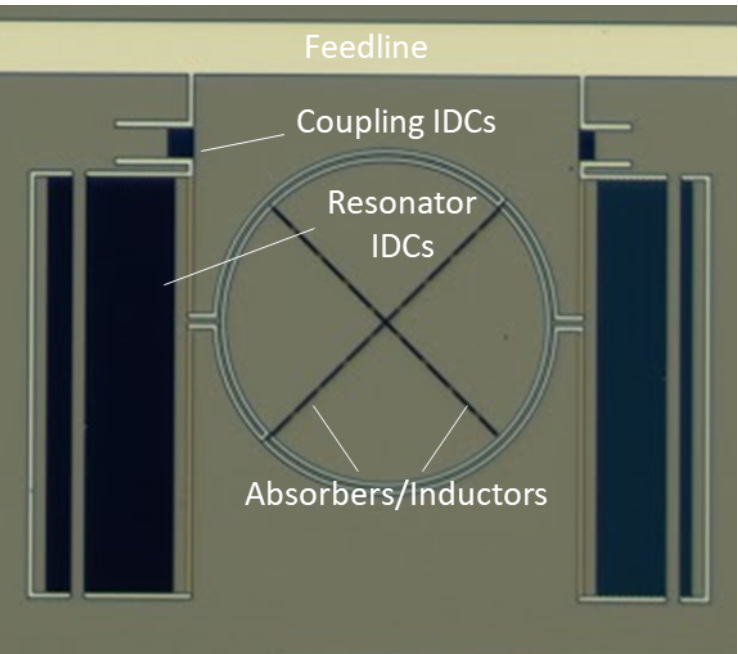}
\caption{A polarized kinetic inductance pixel coupled to a microwave feedline. Each pixel contains two polarized detectors that are coupled to a trimmed interdigitated capacitor. Figure from Duell et al\cite{duell2024superconducting}.}
\label{fig:kid_schematic}
\end{figure}

Kinetic Inductance Detectors (KIDs) are superconducting resonant circuits engineered such that their resonant frequencies shift in response to absorbed radiation \cite{Day_2003}. Each of Mod-Cam's instrument modules employs lumped-element KIDs, which are physically realized as a superconducting inductor that doubles as an absorber, and an interdigitated capacitor. Together they form an $LC$ resonator that is capacitively coupled to a microwave feedline (see Fig.~\ref{fig:kid_schematic}) \cite{Zmuidzinas_2012}. The resonant frequency, $f_0$, is determined by the resonator's total capacitance, $C$, and total inductance, $L_{total}$. The total inductance is the sum of the standard \textbf{geometric inductance} ($L_g$) and the \textbf{kinetic inductance} ($L_k$), which arises from the inertia of the superconducting Cooper pairs:
\begin{equation}
f_0 = \frac{1}{2\pi\sqrt{L_{total}C}} = \frac{1}{2\pi\sqrt{(L_g + L_k)C}}
\end{equation}
The absorption of photons breaks Cooper pairs, which alters the supercurrent density. Since kinetic inductance is sensitive to this density, absorbed power modifies $L_k$ and induces a measurable shift in $f_0$. Similarly, thermal energy from the detector's environment can also break Cooper pairs, creating a sensitivity to the bath temperature known as thermal responsivity. Therefore, by monitoring the complex transmission of a probe signal through the feedline, these frequency shifts can be measured to infer either the incident optical power or fluctuations in the operating temperature \cite{Zmuidzinas_2012}.

% --- EXPERIMENTAL SETUP ---
\section{Experimental Setup}\label{sec:setup}
The measurements described below were performed in Mod-Cam, a single instrument module cryogenic receiver. We briefly overview the Mod-Cam and instrument module design and give a status update here; for more detail, see  \cite{vavagiakis2022ccatprimedesignmodcamreceiver, DuellVavagiakisSPIE}. 

%using the Mod-Cam testbed, a cryogenic instrument platform designed to replicate the operational environment of the Fred Young Submillimeter Telescope (FYST).

\subsection{Mod-Cam Cryogenic and Mechanical Design}

\begin{figure}
\centering
\includegraphics[width=1.0\columnwidth]{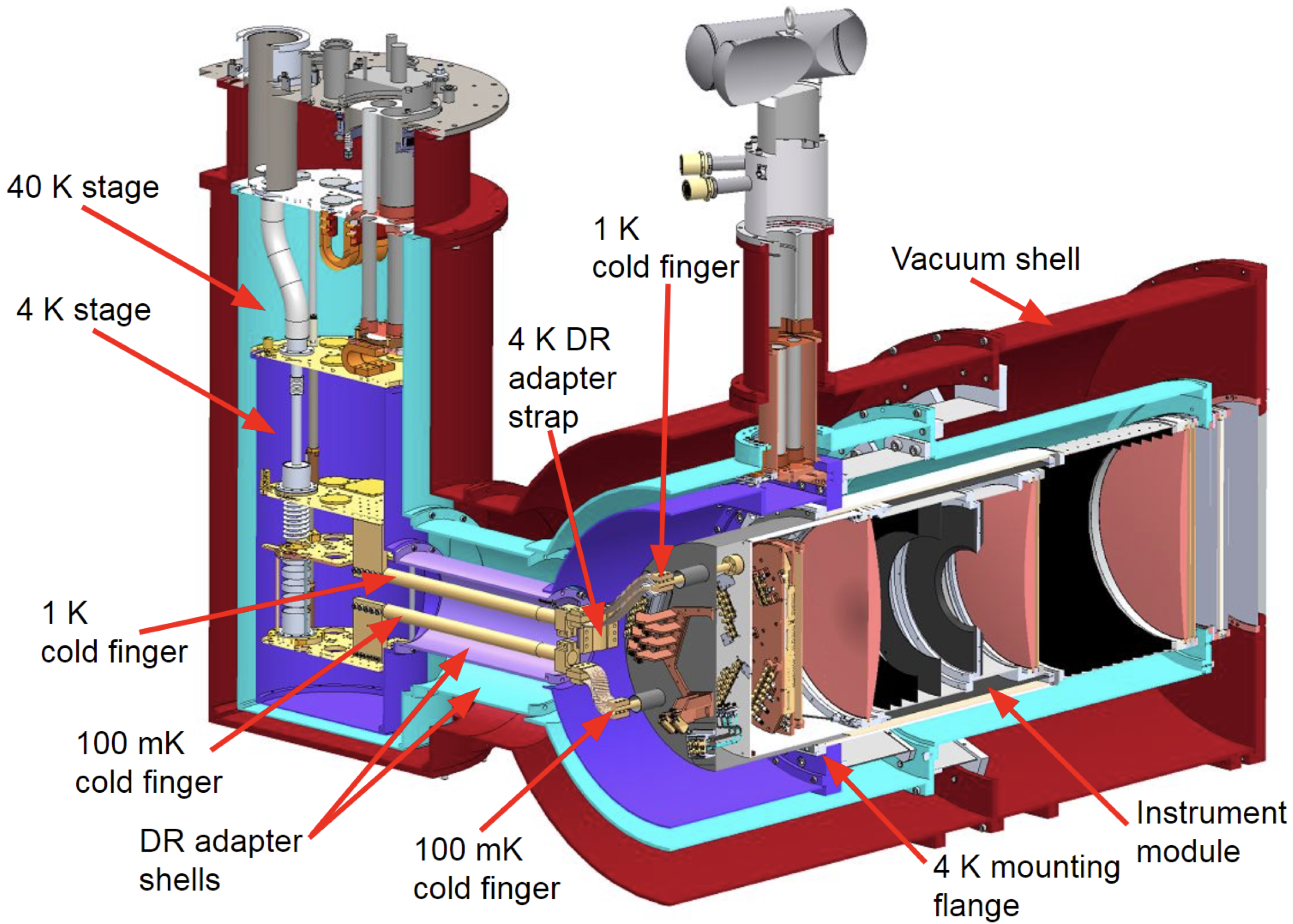}
\caption{A rendered cross-section of Mod-Cam with an instrument module (fig. from \cite{vavagiakis2022ccatprimedesignmodcamreceiver}). Visible stages include the vacuum shell (red), 40 K (light blue), and 4 K (purple). The 1 K and 100 mK stages of the DR are coupled via cold fingers and straps to the instrument module cold fingers, as labeled. The instrument module cold fingers are attached to the 1 K and 100 mK stages of the instrument module, which is shown in cross section and mounted on Mod-Cam's 4 K plate. The KID arrays are mounted on the 100 mK plate of the instrument module.}
\label{fig:modcamxsec}
\end{figure}

Mod-Cam is a 0.9 m diameter, 1.8 m long cryogenic receiver with an off-axis dilution refrigerator (DR) design and 40 K and 4 K stages (Fig. \ref{fig:modcamxsec}), that has been developed and tested at Cornell University. 

Mod-Cam's 6061-T6 Al vacuum shell consists of front and rear shells, front and back plates, a two-piece DR shell, and a DR bottom plate. The front vacuum plate holds a 44 cm hexagonal ultra-high-molecular-weight polyethylene (UHMWPE) vacuum window and optical filters. Mod-Cam's 40 K stage is primarily fabricated of 6063-T5 Al and consists of a front and back shell, front filter plate, back plate, DR shield, DR shield bottom plate, DR shell to main cylinder adapter, and a mounting ring, on which epoxied G-10 tabs hold the 40 K shell assembly off of the vacuum shell. Mod-Cam's 4 K stage, fabricated from 6061-T6 Al, consists of a front plate, shell, rear plate, DR shield, DR shield bottom plate, and DR shell to main cylinder adapter. The 4 K plate is the only mechanical mounting point for the instrument modules, and is supported by and thermally isolated from the 40 K ring by a series of epoxied G-10 tabs. Mod-Cam's 40 and 4 K shells are cooled by a Cryomech PT-420. 

\begin{figure}
\centering
\includegraphics[width=0.7\columnwidth]{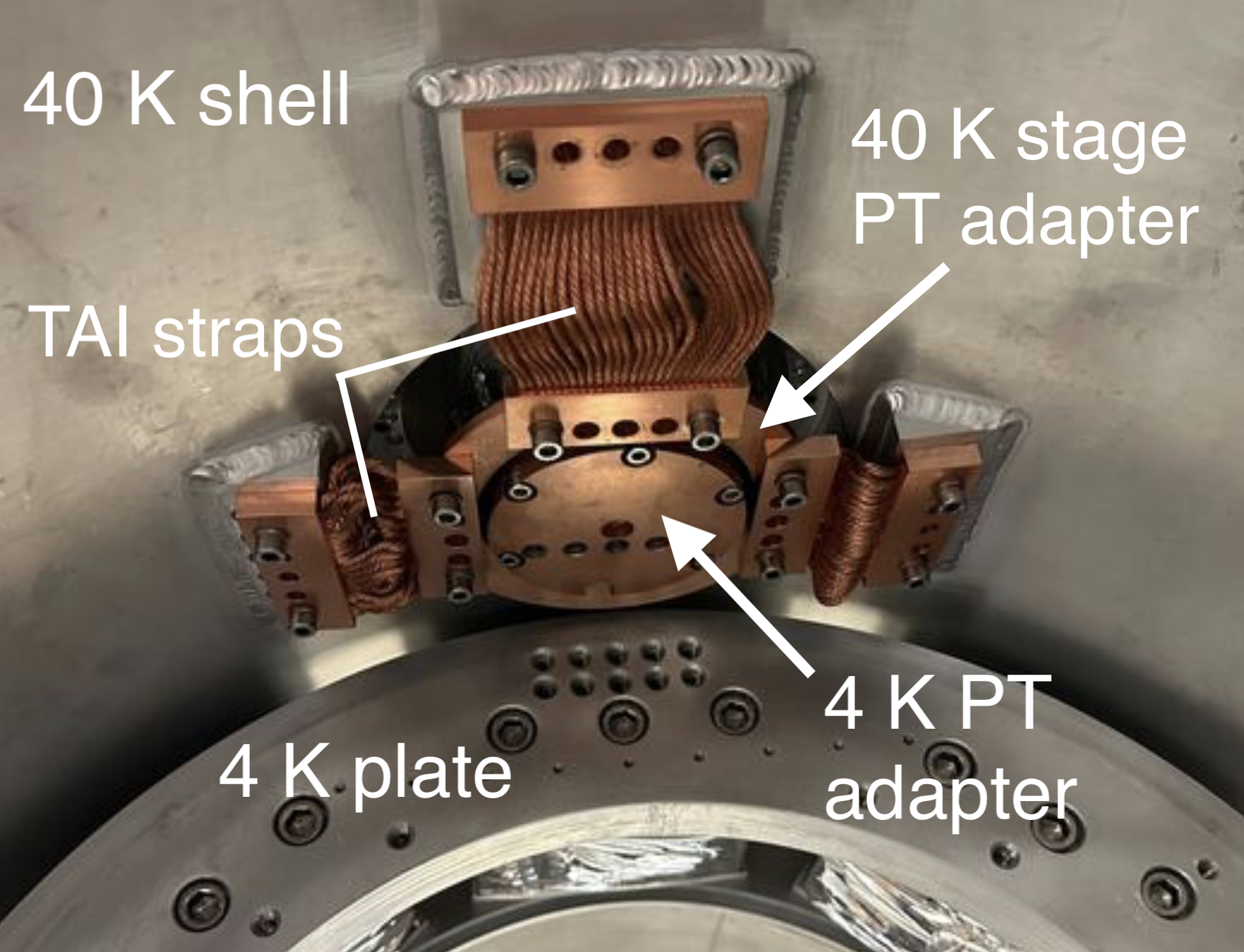}
\caption{The updated Mod-Cam 40 K shell PT-420 interface, including three welded bosses on the Mod-Cam 40 K stage and three OFHC copper TAI straps. This configuration improved the thermal conductivity of the interface, lowering the 40 K front plate base temperature from 72.8K to 55K.}
\label{fig:40KPTstraps}
\end{figure}

To address the upward drift of the 40 K stage temperatures at a rate of 0.2 K/day observed in our previous configuration using homemade clamped OFHC copper ribbon straps and a copper adapter, we upgraded the 40 K thermal interface to use three braided OFHC copper straps from TAI connected to welded bosses on the 40 K shell (Fig. \ref{fig:40KPTstraps}). To improve 1 K and 100 mK base temperatures, the 40 and 4 K stages of the DR are no longer coupled to the main optical access 40 and 4 K Mod-Cam shells. After these changes were made we observed an acceptable \textless 0.1 K/day drift on the 40 K stage. 

\subsection{100 mK Design}

Kinetic inductance detectors must be operated at a temperature significantly below their superconducting critical temperature ($T_c$) to minimize noise from thermally-excited quasiparticles. For the aluminum detectors in this work with a $T_c \approx 1.4$~K \cite{Vaskuri_2025}, the nominal operating temperature is set to 100~mK to ensure this thermal noise is a sub-dominant contribution. To achieve this temperature and the required stability, a Bluefors LD-400 dilution refrigerator's 1\,K stage (still plate) and mixing chamber (MXC) plate are thermally coupled to the 1 K and 100 mK stages of the instrument module (Sec. \ref{sec:instmod}). The still and MXC plates have nominal base temperatures of $\sim$1K and 38\,mK, respectively. The thermal couplings to the DR plates include two gold-plated OFHC copper blocks, connected to two 1.0-inch diameter OFHC copper cold fingers that extend into the main shell space (Fig. \ref{fig:modcamxsec}). The ends of these cold fingers are gold plated, and gold-plated OFHC copper clamps connect the fingers to two braided OFHC copper TAI straps which connect in turn to the instrument module cold fingers (Fig. \ref{fig:modcad}).

\subsection{Instrument Module Design}\label{sec:instmod}

\begin{figure}
\centering
\includegraphics[width=0.85\columnwidth]{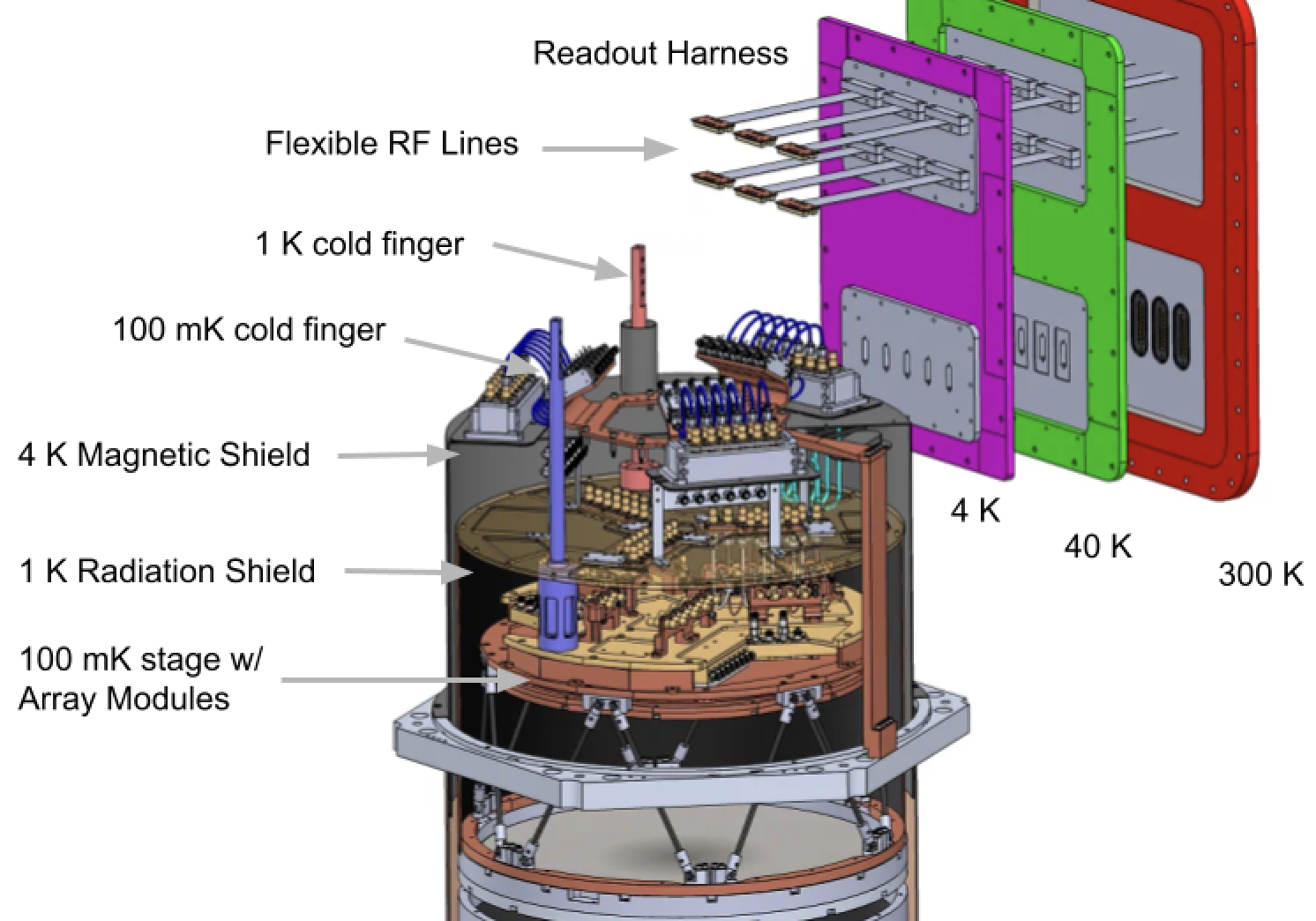}
\caption{Partial cutaway CAD model of the instrument module and cold fingers, along with readout harness, simplified by hiding details such as coaxial cables. The detector arrays are mounted on the 100 mK stage which is offset from the 1 K stage by the carbon fiber truss. The 1 K and 100 mK cold fingers connecting the instrument module stages to the DR cold stages are shown.}
\label{fig:modcad}
\end{figure}

The CAD model of the KID array modules within the instrument module in Mod-Cam is shown in Fig. \ref{fig:modcad}, and was adapted from the Simons Observatory design \cite{Zhu_2021}. The gold-plated OFHC copper instrument module cold fingers are mounted to the 1 K and 100 mK plates of the instrument module. The 100 mK instrument module stage holds the KID array modules and is connected to the 1 K stage by an epoxied carbon fiber truss \cite{vavagiakis2022ccatprimedesignmodcamreceiver}. The portions of the instrument module not shown in Fig. \ref{fig:modcad} include lenses, filters, and other optical elements at and below 4 K, which filter and focus light before it hits the detector arrays. Housed within light-tight enclosures to minimize stray radiation, the detector array temperatures are monitored via two calibrated Lakeshore ROX-102A resistance thermometers. A resistive heater mounted to the MXC plate is used in a PID control loop to servo the detector stage temperature to a stable setpoint, nominally 100 mK.

\subsection{Test Configurations}
All detector performance results in this paper were acquired in a cold load configuration, designed to minimize radiative loading and provide a stable thermal environment for detector characterization. In this setup, the vacuum window was replaced with a blank aluminum cover and a custom cold load assembly was mounted internally to the 40 K stage, replacing the filter and lens stack. This creates a nearly isothermal, dark environment ideal for measuring the system's intrinsic thermal stability and the detectors' thermal responsivity.

For the overall cryogenic characterization in Section V-A, the cryostat was transitioned to an optically open configuration to assess performance in a near field-ready state. This setup involves installing the vacuum window and replacing the previously used metal mesh IR-blocking filters with a room-temperature multi-layer insulation (RT-MLI) style filter made of Zotefoam HD30. The use of Zotefoam for this purpose is a well-established technique in modern cryogenic receivers \cite{Sobrin_2022, Ade_2022}. Our specific configuration used nine 1/8-inch layers and one 1-inch layer of Zotefoam at 300 K, and three 1/8-inch layers at 40 K. The first optical measurements using this setup are presented in a tandem publication by Patel et al. \cite{darshan}.

\begin{figure}
\centering
\includegraphics[width=0.85\columnwidth]{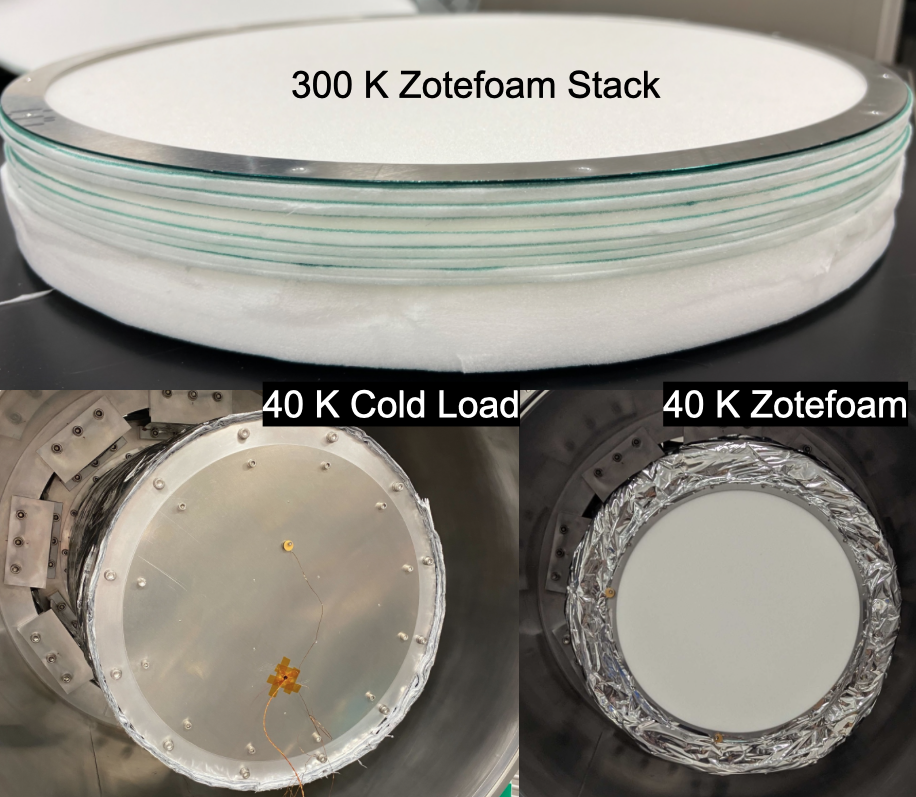}
\caption{Test configurations for Mod-Cam. Detector performance results were obtained using the 40 K cold load (bottom left), installed in place of the optics stack, with the 300 K vacuum window replaced by a blank Al plate. Mod-Cam cryogenic performance was characterized in the ``optically open" configuration with vacuum window along with Zotefoam stacks at the 300 K (top) and 40 K (bottom right) stages.}
\label{fig:configs}
\end{figure}

\begin{comment}
\begin{figure}[!t]
    \centering
    % Top row with the 300 K Zotefoam
    \subfloat[300 K Zotefoam Stack]{\includegraphics[width=0.8\columnwidth]{zotefoam.png}\label{fig:300k_zote}}
    
    % This blank line creates the new row
    
    % Bottom row with the 40 K components side-by-side
    \subfloat[40 K Cold Load]{\includegraphics[width=0.5\columnwidth]{coldload.jpg}\label{fig:cold_load}}
    \hfill
    \subfloat[40 K Zotefoam Stack]{\includegraphics[width=0.45\columnwidth]{40kzote.jpg}\label{fig:40k_zote}}
    
    \caption{Test configurations for Mod-Cam. The detector performance results in this paper were obtained using the 40 K cold load (b), which was installed in place of the 40 K optics stack, while the 300 K vacuum window was replaced with a blank aluminum plate. Mod-Cam cryogenic performance was characterized in the ``optically open" configuration which uses the vacuum window along with Zotefoam stacks at the 300 K (a) and 40 K (c) stages.}
    \label{fig:configs} % A new label for the whole figure
\end{figure}
\end{comment}

\subsection{Readout Electronics Design}
The KID arrays are interrogated by a room-temperature electronics system. A Radio Frequency System-on-Chip  (RFSoC) generates the probe tones and measures the complex transmission ($S_{21}$) through the cryogenic feedline coupled to the detectors. The full readout chain comprises cryogenic coaxial cables, amplifiers, and attenuators, all optimized to maximize the signal-to-noise ratio of the readout tones while minimizing parasitic thermal loading on the cold stages.

\section{Measurement Methodology}
System characterization involved a sequence of measurements to determine the detectors' thermal responsivity and the temperature stability of the cryogenic stage.

\subsection{Thermal Stability Measurement}
To quantify the intrinsic temperature stability of the system, the detector stage was servo-controlled to a constant temperature of 100 mK. The temperature was logged continuously from the nearby thermometer at a 10 Hz sampling rate over an 8-hour period. The root-mean-square (RMS) fluctuation, $\Delta T_{\text{RMS}}$, was then calculated from this time-stream data.

\subsection{Resonator Identification}
Initial identification of each KID's unique resonant frequency is accomplished via a wide-band VNA sweep across the design bandwidth of the array. The built in \texttt{find\_peaks} algorithm in Scipy processes the transmission data to generate a list of approximate resonant frequencies for all detectors.

\subsection{Thermal Responsivity Measurement}
To measure the thermal responsivity, $\mathcal{R}_T = (df_0/f_0)/dT$, the detector stage temperature was deliberately varied. The PID loop controlling the MXC heater was programmed to slowly sweep the detector bath temperature, from 96 mK to 105 mK, with 2 mK step sizes while the temperature was continuously logged. At each stable temperature step, a narrow VNA sweep was executed around each resonator's known frequency found from the previous wide-band VNA sweep. The resulting complex $S_{21}$ data for each resonator are fit in the complex plane to accurately extract the resonant frequency $f_0$.

\begin{figure}[!t]
\centering
\includegraphics[width=1\columnwidth]{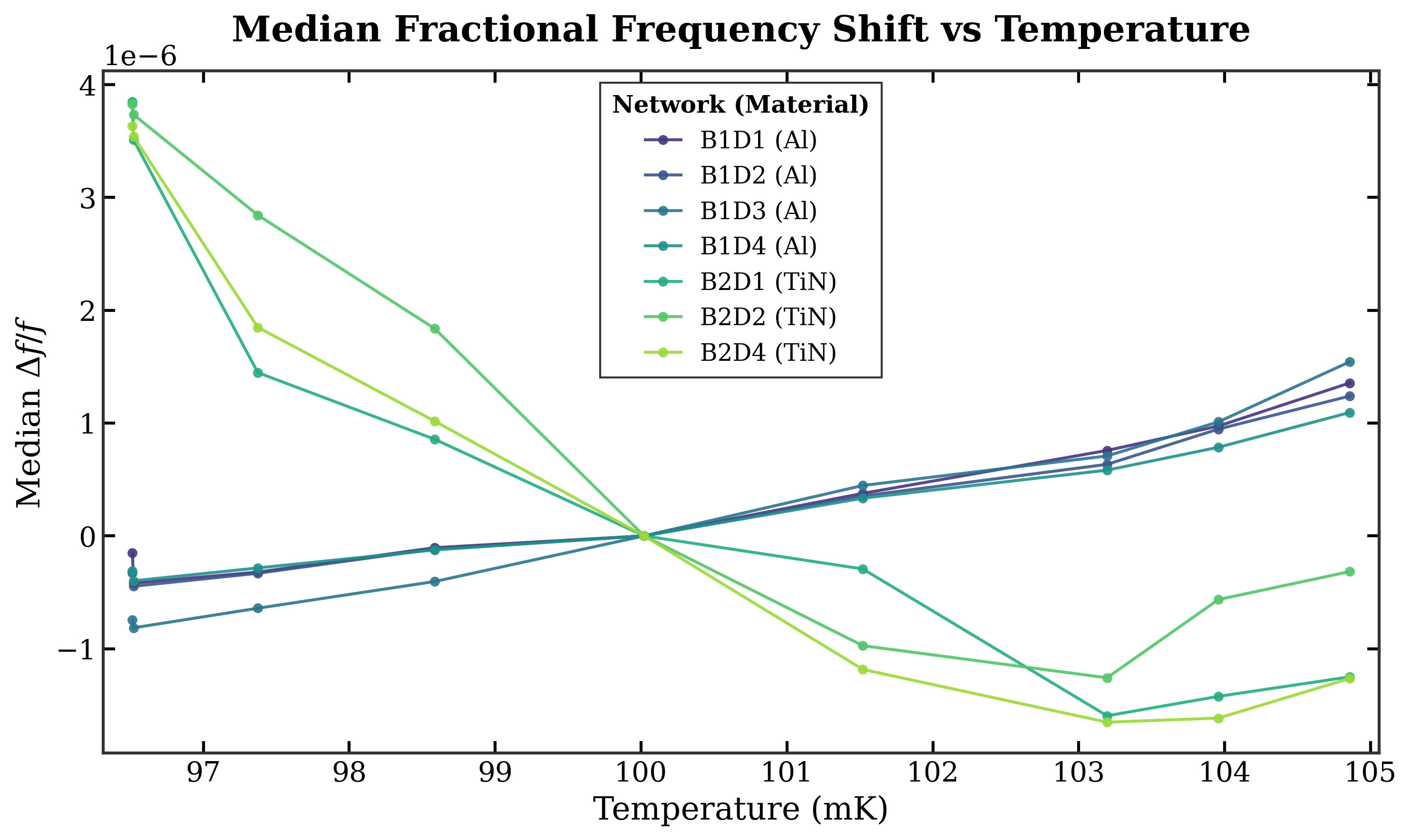}
\caption{Example plot of fractional frequency shift versus detector bath temperature for representative aluminum (Al) and titanium-nitride (TiN) KIDs. The thermal responsivity, $\mathcal{R}_T$, is determined from the slope of a linear fit to these data.}
\label{fig:responsivity_plot}
\end{figure}

The thermal responsivity, $\mathcal{R}_T$, is subsequently determined from the slope of a linear fit to the fractional frequency shift, $\Delta f_0/f_0 = (f_0(T) - f_{0, \text{ref}})/f_{0, \text{ref}}$, versus the change in bath temperature, $\Delta T = T - T_{\text{ref}}$. An example of this measurement is presented in Figure \ref{fig:responsivity_plot}.

% --- RESULTS AND ANALYSIS ---
\section{Results and Analysis}
The characterization data yield an assessment of the system's thermal performance and its impact on detector sensitivity. The methodology for this assessment is to first empirically measuring the detectors' intrinsic thermal responsivity ($\mathcal{R}_T$) and the thermal stability of the focal plane ($\Delta T_{\text{RMS}}$). These results are then combined to calculate the fractional frequency noise induced by the thermal environment. Finally, this instrumental noise is translated into an equivalent optical power fluctuation ($\Delta P_{\text{equiv}}$) by referencing optical responsivity values from prior work \cite{Vaskuri_2025}, which allows for a direct comparison to the expected photon noise floor.

\subsection{Mod-Cam Thermal Performance}

\begin{figure}
\centering
% User should replace 'fig1.png' with an actual file.
\includegraphics[width=1\columnwidth]{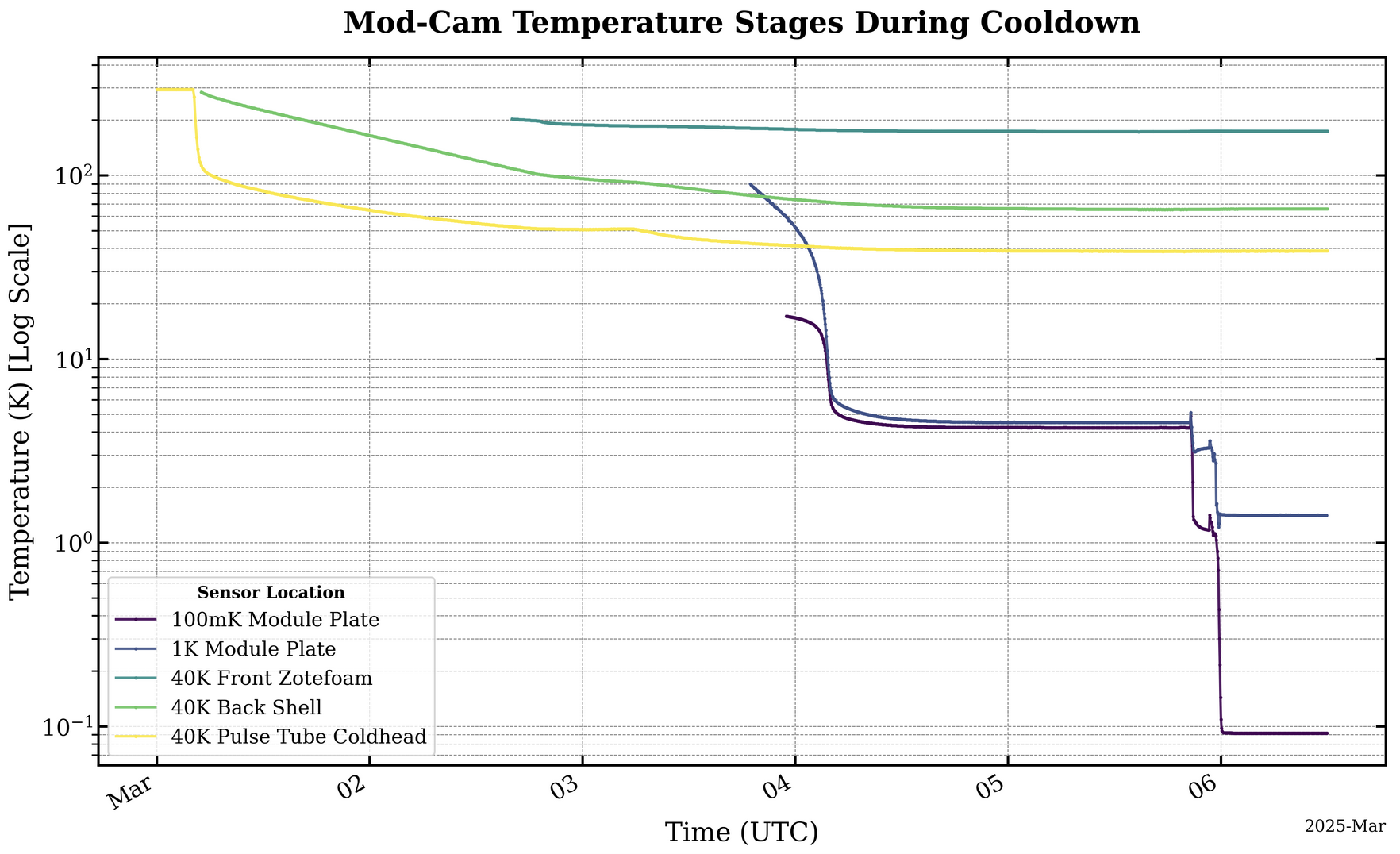}
\caption{Cooldown curve for Mod-Cam in the ``cold load" configuration. Base temperature on the 100 mK stage is achieved approximately 5 days after the start of the cooldown.}
\label{fig:logcooldownplot}
\end{figure}

The thermal performance of the Mod-Cam cryostat was previously validated in a dark configuration without an instrument module \cite{vavagiakis2022ccatprimedesignmodcamreceiver}. Now we present the thermal performance in the ``optically open" configuration described in Sec. \ref{sec:setup}, with the 280 GHz instrument module installed. In this configuration, it takes 1.5 days to evacuate the receiver to below $10^{-3}$ mbar, after which the PTs are turned on. It then takes 3 days to cool to 4 K, after which the DR is condensed. Base temperature for the 100 mK stage is achieved in 1 hour after the start of condensing. A cooldown curve for this data taking run is shown in Fig. \ref{fig:logcooldownplot}. The 40 and 4 K stage cooldown curve agrees with the cooldown model in \cite{gascard2024thermalmechanicalstudyparametrised}. Base temperatures and estimated loading per stage are reported in Table \ref{tab:basetemps}. Gradients between the DR stages and instrument module plates were 0.3 K for the 1 K stage and 40 mK for the 100 mK stage.

\begin{table}[!t]
\caption{Base Temperature Comparison of Mod-Cam in Cold Load vs. Optically Open Configurations}
\label{tab:basetemps}
\centering
\begin{tabular}{@{}lcc@{}}
\toprule
& \textbf{Base Temp.} & \textbf{Base Temp.} \\
& \textbf{(Cold Load) [K]} & \textbf{(Optically Open) [K]} \\
\midrule
40 K Front Shell & 57 & 173K \\
4 K LNA Heat Sink & 7.1 & 9.6 \\
Module 1 K Plate & 1.3 & 1.4 \\
Module 0.1 K Plate & 0.097 & 0.101 \\
\bottomrule
\end{tabular}
\end{table}

\subsection{Thermal Stability of the 100 mK Stage}
\begin{figure}[!t]
\centering
% User should replace 'fig1.png' with an actual file.
\includegraphics[width=1\columnwidth]{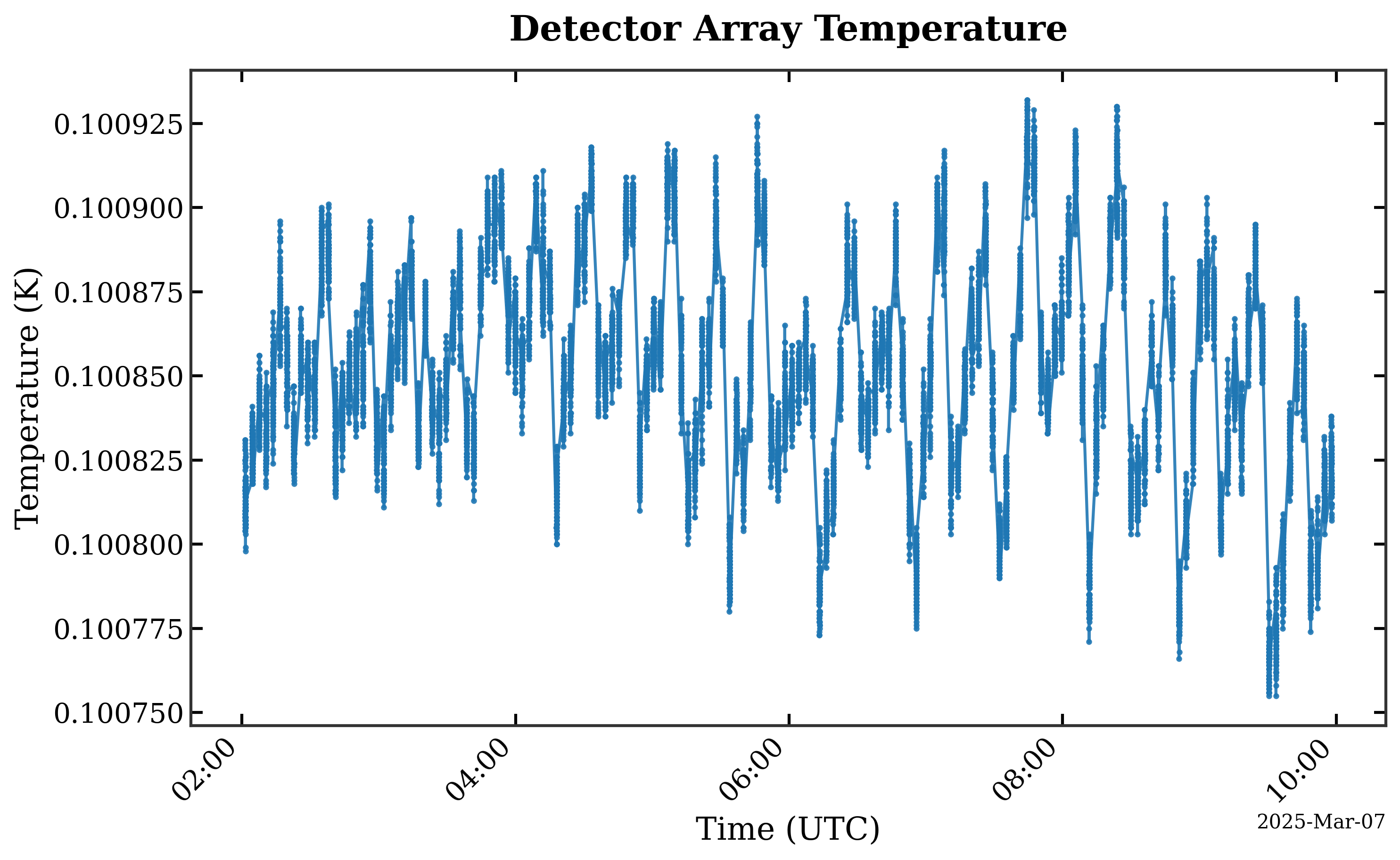}
\caption{A 30-minute segment of the temperature time-stream from the detector array thermometer while servo-controlled at 100 mK. The RMS fluctuation over the full 8-hour measurement period was found to be $3.2 \times 10^{-5} \text{ K}$.}
\label{fig:temp_timestream}
\end{figure}

The following detector stability measurements were performed in the cold load configuration to isolate the system from external radiative power.
The detector array temperature was logged while being servo-controlled at 100 mK. Figure \ref{fig:temp_timestream} shows a representative segment of the time-stream data. Analysis of the full 8-hour dataset yielded an RMS temperature fluctuation of:
$$
\Delta T_{\text{RMS}} = 3.2 \times 10^{-5} \, \text{K}
$$
This high degree of stability demonstrates the efficacy of the cryogenic design and PID control loop.

\subsection{KID Thermal Responsivity}
Thermal responsivity measurements were performed for 2409 aluminum detectors and 1886 titanium-nitride detectors on the 280\,GHz array. The median responsivity values were found to be:
\begin{itemize}
    \item \textbf{Aluminum (Al):} $\mathcal{R}_{T, \text{Al}} = \SI{1.995e-4}{\per\kelvin} \pm \SI{3.11e-3}{\per\kelvin}$
    \item \textbf{Titanium Nitride (TiN):} $\mathcal{R}_{T, \text{TiN}} = \SI{-6.081e-4}{\per\kelvin} \pm \SI{1.29e-2}{\per\kelvin}$
\end{itemize}
The TiN detectors exhibit a significantly higher magnitude of thermal responsivity, a consequence of both a higher kinetic inductance fraction and a lower superconducting gap energy for these specific devices relative to the Al detectors. The opposite signs of the responsivity are also expected. The negative responsivity of TiN follows the standard theory of microwave conductivity in superconductors \cite{Mattis_1958}, where a rising temperature breaks Cooper pairs, increasing kinetic inductance and thus decreasing the resonant frequency. Conversely, the positive responsivity in the aluminum KIDs is dominated by a competing nonlinear effect from interactions with two-level system (TLS) defects \cite{Gao_2008_TLS}. 

\subsection{Derivation of Equivalent Optical Power Fluctuation}
A key analysis step translates the measured temperature noise into an equivalent optical power fluctuation, enabling a direct comparison between the instrumental noise floor and the fundamental photon noise. First, the RMS fractional frequency noise induced by thermal fluctuations, $(\Delta f/f_0)_{\text{RMS, thermal}}$, is calculated:
\begin{equation}
\left( \frac{\Delta f}{f_0} \right)_{\text{RMS, thermal}} = |\mathcal{R}_T| \times \Delta T_{\text{RMS}}
\label{eq:thermal_freq_noise}
\end{equation}
Next, we define the optical responsivity, $R_{\text{opt}}$, which relates a change in incident optical power, $\delta P_{\text{opt}}$, to the resulting fractional frequency shift:
\begin{equation}
S_{\text{opt}}(P_{\text{opt}}) = \frac{\delta f / f_0}{\delta P_{\text{opt}}}
\end{equation}
The equivalent optical power fluctuation, $\Delta P_{\text{equiv}}$, is defined as the optical power fluctuation that would produce the same RMS frequency shift as the thermal bath instability,
\begin{equation}
\Delta P_{\text{equiv}} = \frac{(\Delta f/f_0)_{\text{RMS, thermal}}}{|S_{\text{opt}}(P_{\text{opt}})|}.
\label{eq:delta_p_equiv}
\end{equation}

\subsection{Calculation and Comparison to Noise Benchmarks}
To calculate $\Delta P_{\text{equiv}}$, we utilize optical responsivity values, $R_{\text{opt}}$, from related work characterizing similar detectors \cite{Vaskuri_2025}, assuming a typical background photon power for FYST in the range of 5--9 pW.

\subsubsection{Calculations for aluminum (Al)}
Using $\mathcal{R}_{T, \text{Al}} = \SI{1.995e-4}{\per\kelvin}$ and $\Delta T_{\text{RMS}} = \SI{3.2e-5}{\kelvin}$:
$$
\left( \frac{\Delta f}{f_0} \right)_{\text{RMS, Al}} \approx 6.38 \times 10^{-9}
$$
Using $|S_{\text{opt,Al}}(5\,\text{pW})| \approx \SI{3.18e7}{\per\watt}$ \cite{Vaskuri_2025}:
\begin{align}
\Delta P_{\text{equiv, Al}}(5\,\text{pW}) &= \frac{6.38 \times 10^{-9}}{\SI{3.18e7}{\per\watt}} \nonumber \\
&\approx 2.01 \times 10^{-16} \, \text{W} \approx \mathbf{0.20} \, \textbf{fW}
\end{align}

\subsubsection{Calculations for Titanium Nitride (TiN)}
Using $\mathcal{R}_{T, \text{TiN}} = \SI{-6.081e-4}{\per\kelvin}$:
$$
\left( \frac{\Delta f}{f_0} \right)_{\text{RMS, TiN}} \approx 1.95 \times 10^{-8}
$$
Using $|S_{\text{opt,TiN}}(5\,\text{pW})| \approx \SI{1.70e8}{\per\watt}$ \cite{Vaskuri_2025}:
\begin{align}
\Delta P_{\text{equiv, TiN}}(5\,\text{pW}) &= \frac{1.95 \times 10^{-8}}{\SI{1.70e8}{\per\watt}} \nonumber \\
&\approx 1.15 \times 10^{-16} \, \text{W} \approx \mathbf{0.11} \, \textbf{fW}
\end{align}
These calculations were repeated for a 9 pW optical load, with results summarized in Table \ref{tab:results_summary}.

\begin{table}[!t]
\caption{Summary of Thermally-Induced Equivalent Power Noise}
\label{tab:results_summary}
\centering
\begin{tabular}{@{}lcc@{}}
\toprule
\textbf{Detector Material} & \multicolumn{2}{c}{\textbf{$\Delta P_{\text{equiv}}$ [\si{\femto\watt}]}} \\
\cmidrule(l){2-3}
& at 5 pW Load & at 9 pW Load \\
\midrule
aluminum (Al) & 0.20 & 0.27 \\
Titanium Nitride (TiN) & 0.11 & 0.13 \\
\bottomrule
\end{tabular}
\end{table}

For a typical 5~pW optical load, the $\Delta P_{\text{equiv}}$ of 0.20 fW for the aluminum detectors corresponds to a fractional noise contribution of only 0.0040\% ($\Delta P_{\text{equiv}} / P_{\text{opt}}$). The result is even better for the titanium nitride detectors, where the 0.11 fW of noise contributes just 0.0023\%. This demonstrates that the noise from cryogenic thermal fluctuations is insignificant compared to the expected background signal, and will have a negligible impact on the instrument's sensitivity.

\section{Discussion}
The results presented demonstrate that the Mod-Cam cryogenic system performs to specification. The achieved temperature stability of $\Delta T_{\text{RMS}} = \SI{3.2e-5}{\kelvin}$ at the 100 mK stage is sufficient to ensure that instrument sensitivity is not limited by thermal fluctuations. The resultant equivalent optical power noise corresponds to a fractional impact on a nominal 5~pW background signal of only 0.0040\% for the aluminum detectors and 0.0023\% for the titanium nitride detectors. This result confirms that thermal fluctuations are expected to have negligible impact on the instrument white noise level, a primary design objective for any high-sensitivity astronomical instrument.

The small residual temperature fluctuations likely arise from a combination of factors, including microphonics from the pulse tube cryocooler, imperfect PID parameters and minute fluctuations in the DR cooling power. As the analysis shows, however, the impact of these residual fluctuations on the noise budget is negligible. Future work will involve performing these characterizations under direct optical loading from a cryogenic blackbody source to measure the optical responsivity, $R_{\text{opt}}$, directly and provide a complete portrait of detector performance in its final operational state.

\section{Conclusion}
We have performed a detailed thermal characterization of the 280 GHz instrument module arrays for the Mod-Cam receiver. The thermal responsivity of both Al and TiN detectors was measured and the temperature stability of the cryogenic detector stage was quantified. The principal result is that the measured RMS temperature fluctuation of $\SI{3.2e-5}{\kelvin}$ at 100 mK contributes a negligible noise component to the detector readout. The fractional impact of the equivalent power from thermal fluctuations has negligible contribution to the power of incident photons. This result validates the cryogenic design and thermal control systems for Mod-Cam and provides high confidence that the instrument will achieve its background-limited sensitivity goals when deployed on the Fred Young Submillimeter Telescope.

\section*{Acknowledgments}
The CCAT project, FYST and Prime-Cam instrument have been supported by generous contributions from the Fred M. Young, Jr. Charitable Trust, Cornell University, and the Canada Foundation for Innovation and the Provinces of Ontario, Alberta, and British Columbia. The construction of the FYST telescope was supported by the Gro{\ss}ger{\"a}te-Programm of the German Science Foundation (Deutsche Forschungsgemeinschaft, DFG) under grant INST 216/733-1 FUGG, as well as funding from Universit{\"a}t zu K{\"o}ln, Universit{\"a}t Bonn and the Max Planck Institut f{\"u}r Astrophysik, Garching. The construction of the 350 GHz instrument module for Prime-Cam is supported by NSF grant AST-2117631. S. Chapman acknowledges support from NSERC and CFI. S. Walker acknowledges support from the National Science Foundation under Award No. 2503181.

\balance % This should be the last command before the bibliography

\printbibliography

\end{document}